\documentclass[twocolumn,times]{aastex62}
\usepackage{rotating}
\usepackage{graphicx}
\usepackage{gensymb}
\usepackage{appendix}
\usepackage{float}
\usepackage[normalem]{ulem}     

\newcommand{\angstrom}{\mbox{\normalfont\AA}}

\graphicspath{{./}{figures/}}

\received{XXXX, 2018}
\revised{XXXX, 2018}
\accepted{\today}
\submitjournal{ApJ}

\shorttitle{Transient IFIP}
\shortauthors{Baker et al.}

\begin{document}

\title{Transient Inverse-FIP Plasma Composition Evolution within a Confined Solar Flare}

\correspondingauthor{Deborah Baker}
\email{deborah.baker@ucl.ac.uk}

\author[0000-0002-0665-2355]{Deborah Baker}
\affil{University College London, Mullard Space Science Laboratory, Holmbury St. Mary, Dorking, Surrey, RH5 6NT, UK}

\author[0000-0002-2943-5978]{Lidia van Driel-Gesztelyi}
\affiliation{University College London, Mullard Space Science Laboratory, Holmbury St. Mary, Dorking, Surrey, RH5 6NT, UK}
\affiliation{LESIA, Observatoire de Paris, Universit\'e PSL, CNRS, Sorbonne Universit\'e, Univ. Paris Diderot, Sorbonne Paris Cit\'e, 5 place Jules Janssen, 92195 Meudon, France}
\affiliation{Konkoly Observatory, Research Centre for Astronomy and Earth Sciences, Hungarian Academy of Sciences, Konkoly Thege \'ut 15-17., H-1121, Budapest, Hungary}

\author[0000-0002-2189-9313]{David H. Brooks}
\affiliation{College of Science, George Mason University, 4400 University Drive, Fairfax, VA 22030, USA}

\author[0000-0001-7809-0067]{Gherardo Valori}
\affiliation{University College London, Mullard Space Science Laboratory, Holmbury St. Mary, Dorking, Surrey, RH5 6NT, UK}

\author[0000-0001-7927-9291]{Alexander W. James}
\affiliation{University College London, Mullard Space Science Laboratory, Holmbury St. Mary, Dorking, Surrey, RH5 6NT, UK}

\author[0000-0002-3362-7040]{J. Martin Laming}
\affiliation{Space Science Division, Naval Research Laboratory, Code 7684, Washington, DC 20375, USA}

\author[0000-0003-3137-0277]{David M. Long}
\affiliation{University College London, Mullard Space Science Laboratory, Holmbury St. Mary, Dorking, Surrey, RH5 6NT, UK}

\author[0000-0001-8215-6532]{Pascal D\'emoulin}
\affiliation{LESIA, Observatoire de Paris, Universit\'e PSL, CNRS, Sorbonne Universit\'e, Univ. Paris Diderot, Sorbonne Paris Cit\'e, 5 place Jules Janssen, 92195 Meudon, France}

\author[0000-0002-0053-4876]{Lucie M. Green}
\affiliation{University College London, Mullard Space Science Laboratory, Holmbury St. Mary, Dorking, Surrey, RH5 6NT, UK}

\author[0000-0001-9346-8179]{Sarah A. Matthews}
\affiliation{University College London, Mullard Space Science Laboratory, Holmbury St. Mary, Dorking, Surrey, RH5 6NT, UK}

\author[0000-0001-9570-3558]{Katalin Ol\'{a}h}
\affiliation{Konkoly Observatory of the Hungarian Academy of Sciences, Budapest, Hungary}

\author[0000-0001-5160-307X]{Zsolt K\H{o}v\'{a}ri}
\affiliation{Konkoly Observatory of the Hungarian Academy of Sciences, Budapest, Hungary}

\begin{abstract}
Understanding elemental abundance variations in the solar corona provides an insight into how matter and energy flow from the chromosphere into the heliosphere. 
Observed variations depend on the first ionization potential (FIP) of the main elements of the Sun's atmosphere.  
High-FIP elements ($>$10 eV) maintain photospheric abundances in the corona, whereas low-FIP elements have enhanced abundances. 
Conversely, inverse FIP (IFIP) refers to the enhancement of high-FIP or depletion of low-FIP elements. 
We use spatially resolved spectroscopic observations, specifically the Ar {\sc xiv}/Ca {\sc xiv} intensity ratio, from \emph{Hinode's} Extreme-ultraviolet Imaging Spectrometer to investigate the distribution and evolution of plasma composition within two confined flares in a newly emerging, highly sheared active region.  
During the decay phase of the first flare, patches above the flare ribbons evolve from the FIP to the IFIP effect, while the flaring loop tops show a stronger FIP effect.  
The patch and loop compositions then evolve toward the pre-flare basal state. 
We propose an explanation of how flaring in strands of highly sheared emerging magnetic fields can lead to flare-modulated IFIP plasma composition over coalescing umbrae which are crossed by flare ribbons. 
Subsurface reconnection between the coalescing umbrae leads to the depletion of low-FIP elements as a result of an increased wave flux from below. 
This material is evaporated when the flare ribbons cross the umbrae. 
Our results are consistent with the ponderomotive fractionation model \citep{laming15} for the creation of IFIP-biased plasma. 
\end{abstract}

\keywords{Sun: abundances - Sun: corona - Sun: magnetic fields}

\section{Introduction} \label{sec:intro}

Elemental abundance variations are tracers of physical processes throughout the Universe, with the cosmic reference standard being the solar chemical composition.
Understanding how the Sun's chemical composition varies in time and space provides an insight into how mass and energy flow from the Sun's chromosphere into the heliosphere and in turn, from the chromospheres of solar-like stars into their astrospheres \citep[e.g.,][]{testa10,testa15,laming15}.
 
From over 50 years of spectroscopic observations we know that the solar corona has different elemental composition from that of the photosphere \citep{pottasch63,meyer85a,meyer85b}.
In the corona, elements of low first ionization potential (FIP; $\leq$ 10 eV) such as Fe, Si, Mg, and Ca are enhanced by a factor of two to four compared to their photospheric abundances. 
This is known as the FIP effect and it is typically expressed in terms of the FIP bias parameter which is the ratio of an element's coronal and photospheric abundances.
The enhancement of low-FIP elements varies depending on the coronal structure, e.g., coronal holes show little plasma fractionation 
\citep[FIP bias $\sim$1 or photospheric composition;][]{feldman93,feldman98,doschek98,brooks11,baker13}, quiet-Sun regions typically have a FIP bias of 1.5--2 \citep{feldman93,doschek98,warren99,baker13,ko16,baker18}, whereas active regions (ARs) consist of highly fractionated plasma with FIP bias of 3--4 \citep{feldman92,widing95,sheeley96,brooks11,brooks12,baker13,delzanna14,baker15,brooks15}.
(See reviews by e.g., \cite{feldman03,schmelz12,laming15} and Chapter 14 of \cite{delzanna18}). 

Unresolved Sun-as-a-star observations using full disk integrated spectra show the solar FIP effect of the active Sun.  
For temperatures greater than $\sim$1 MK, \cite{laming95} found low-FIP elements were enhanced by a factor of 3--4 which is in line with the FIP bias value for the element Fe deduced by \cite{schonfeld15}.
The FIP effect is reduced at lower temperatures \citep{laming95}.
Recent work of \cite{brooks17,brooks18a} based on Sun-as-a-star spectra obtained by the Solar Dynamics Observatory (SDO) Extreme-Ultraviolet Variability Experiment (EVE) demonstrated that the variation of coronal composition is highly correlated with the solar cycle.
 
The coronae of solar-like stars exhibit a varying degree of the solar-like FIP effect in their X-ray spectra whereas more active cool dwarf stars show an inverse FIP (IFIP) effect.  In surveys of M dwarf stars and active binaries, low-FIP elements are under-abundant relative to high-FIP elements e.g., O, Ne, Ar \citep[\emph{e.g.,}][]{brinkman01,telleschi05,robrade05,argiroffi05,robrade06,wood06,liefke08,wood12}.
\cite{wood10} firmly established the dependence of the (I)FIP effects on F to K spectral type stars with X-ray luminosity $<$10$^{29}$ ergs s$^{-1}$.
The solar FIP effect decreases from G to early K-type stars and becomes zero at $\sim$ K5 then reverses to the IFIP effect for later K stars and M dwarfs.
\cite{wood18} extended the FIP bias--spectral type relationship to include stars of earlier spectral types A and F.
It is worth noting that the observed composition of stellar coronae can be more complex than having either a straightforward FIP or IFIP effect. 
For example, \cite{peretz15} found that all elements were consistently depleted in the coronae of six main sequence stars of spectral type F7--K1 compared to their respective photospheres, whether compared to solar abundances or the individual stellar abundances.

Surveys of plasma composition in solar flares show notable variability in elemental abundances.
\cite{warren14a} measured absolute abundances for 21 M9.3 to X6.9 class flares using \emph{SDO}/EVE spectra.
The mean FIP bias in their sample is close to photospheric composition which is consistent with the earlier results of \cite{veck81,feldman90,mckenzie92} and more recently with \cite{delzanna13b}.
However, in large samples of solar flares observed by multiple instruments, low-FIP element abundances are enhanced by a factor of $\sim$2--3, depending on the emission lines and instruments used to determine the FIP bias, the atomic data used at the time of the measurements, and temperature effects \citep[e.g.,][]{doschek85,sterling93,bentley97,fludra99,phillips12,dennis15,sylwester15}.
Spatially resolved flare observations from \emph{Hinode}/EIS provide clues to understanding the abundance variability observed in solar flares.
\cite{doschek18} show that the FIP bias varies from coronal composition in the post-flare loops to photospheric composition in the loop footpoints for an X8.3 flare and \cite{warren18} find coronal abundances in the current sheet of the same limb flare.
At least for the solar case, it is possible that the spatially unresolved observations \citep[e.g.,][]{warren14a} show little FIP effect in flares because photospheric abundances are dominant and the coronal composition of specific features such as post-flare loops is only evident in spatially resolved images.

Studies of abundance changes in stellar flares have been mainly limited to large flares on active stars i.e., to stars with IFIP composition.
Using high resolution X-ray spectroscopy, \cite{nordon07,nordon08} analyzed abundance variations during 14 large flares observed by the \emph{XMM-Newton} and \emph{Chandra} observatories.
In 7 of 14 flares, they found a trend of enhanced low-FIP elements tending towards photospheric composition during the flares for stars with IFIP-biased quiescent coronae.
The opposite trend was observed in two stars with FIP--bias dominated plasma comprising their coronae and five stars showed no effect during flaring.
Their results are consistent with case studies of solar-like stars \citep[e.g.,][]{testa07}, active M dwarf EV Lac \citep{laming09b}, active M dwarf CN Leo \citep{liefke10}, and RS CVn binary stars \citep[e.g.,][]{gudel99,audard01,audard03}.
Indeed, our knowledge is presently limited for flares in stars because of the difficulties in determining abundance variations in stellar coronae; see the extensive discussion in \cite{testa10}.

Any theoretical framework for a fractionation mechanism must be able to account for all of the solar and stellar observations including the IFIP effect.
Early models based on, for example, thermal diffusion and Coulomb drag could reproduce some observational aspects of the FIP effect but not those of the IFIP effect \citep[][ and references therein]{laming15}.
To date, only the ponderomotive force model is able to produce  IFIP composition. 
The Laming model invokes the ponderomotive force acting only on ions to separate ions from neutrals in the chromosphere of the Sun and other stars \citep{laming12,laming15,laming17}. 
Fractionation takes place in the chromosphere at temperatures where low-FIP elements are mainly ionized, but high-FIP elements remain neutral. 
The ponderomotive force arises as Alfv\'en waves reflect from or refract at the high density gradient in the chromosphere of the Sun.
Alfv\'en waves originating in the corona cause FIP fractionation at the chromospheric level at loop footpoints which are magnetically connected to reconnection sites of flares or nanoflares  \citep{laming17}.
Conversely, upwardly directed photospheric acoustic waves may mode convert to fast mode waves as the plasma transitions from $\beta$ $>$1 to $\beta$ $<$1, where plasma $\beta$ is the ratio of plasma pressure and magnetic pressure. 
The IFIP effect arises as these new fast mode waves refract in the chromosphere back down again, producing a downward directed ponderomotive acceleration \citep{laming15}.

Recently, \cite{doschek15} reported the first observations of the IFIP effect on the Sun.
Highly localized regions of IFIP were captured by \emph{Hinode}/EIS \citep{culhane07} near large sunspots during flares \citep{doschek15,doschek16,doschek17}.
In the most extreme case, the IFIP effect locally exceeded that of the integrated value of M dwarf stars of spectral type $\sim$M5 at the extreme end of the FIP bias-spectral type relationship of \cite{wood10,laming15,wood18}. 

In this paper, we present a detailed spectroscopic study of the evolution of plasma composition in the very active AR 11429.
We find that patches of IFIP bias plasma appear and disappear within the FIP-bias dominated composition of the AR during a confined flare, while IFIP--bias plasma was not observed in another confined flare nine hours later.
We discuss how the relatively rare magnetic field configuration of the AR and its evolution relate to the spatial locations and temporal evolution of IFIP patches and we consider whether flaring plays a role in creating the anomalous plasma composition or simply reveals the already present IFIP plasma. 

\section{Overview of AR 11429}\label{sec:overview}
\subsection{Coronal Activity}\label{activity}
AR 11429 appeared at the NE limb on 2012 March 3.
During its disk transit, AR 11429 was the source region of three X-class, 14 M-class, 32 C-class flares and four coronal mass ejections (CMEs), making it one of the most flare and CME productive active regions of Solar Cycle 24.
The high activity level and rare magnetic field configuration have inspired a large number of studies of the active region  \citep[\emph{e.g.}][]{petrie12,donea13,sun15,chintzoglou15,syntelis16,patsourakos16,polito17}.
In this study, we focus on two confined M-class flares observed by \emph{Hinode}/EIS on March 6.
They are identified as FL1 and FL2 in the \emph{GOES} 1--8 $\angstrom$ X-ray flux curve in Figure \ref{fig:multi}. 
The first flare starts at 12:23 UT, peaks at 12:37 UT, and decays to background flux levels by $\sim$13:45 UT and the second flare begins at 21:04 UT, peaks at 21:12 UT and quickly decays by $\sim$21:25 UT.
FL1 and FL2 flare classifications are M2.2 and M1.4, respectively.
\bigskip
\bigskip
\subsection{Coronal Evolution}\label{sec:corona}

\begin{figure}
\includegraphics[scale = 0.43]{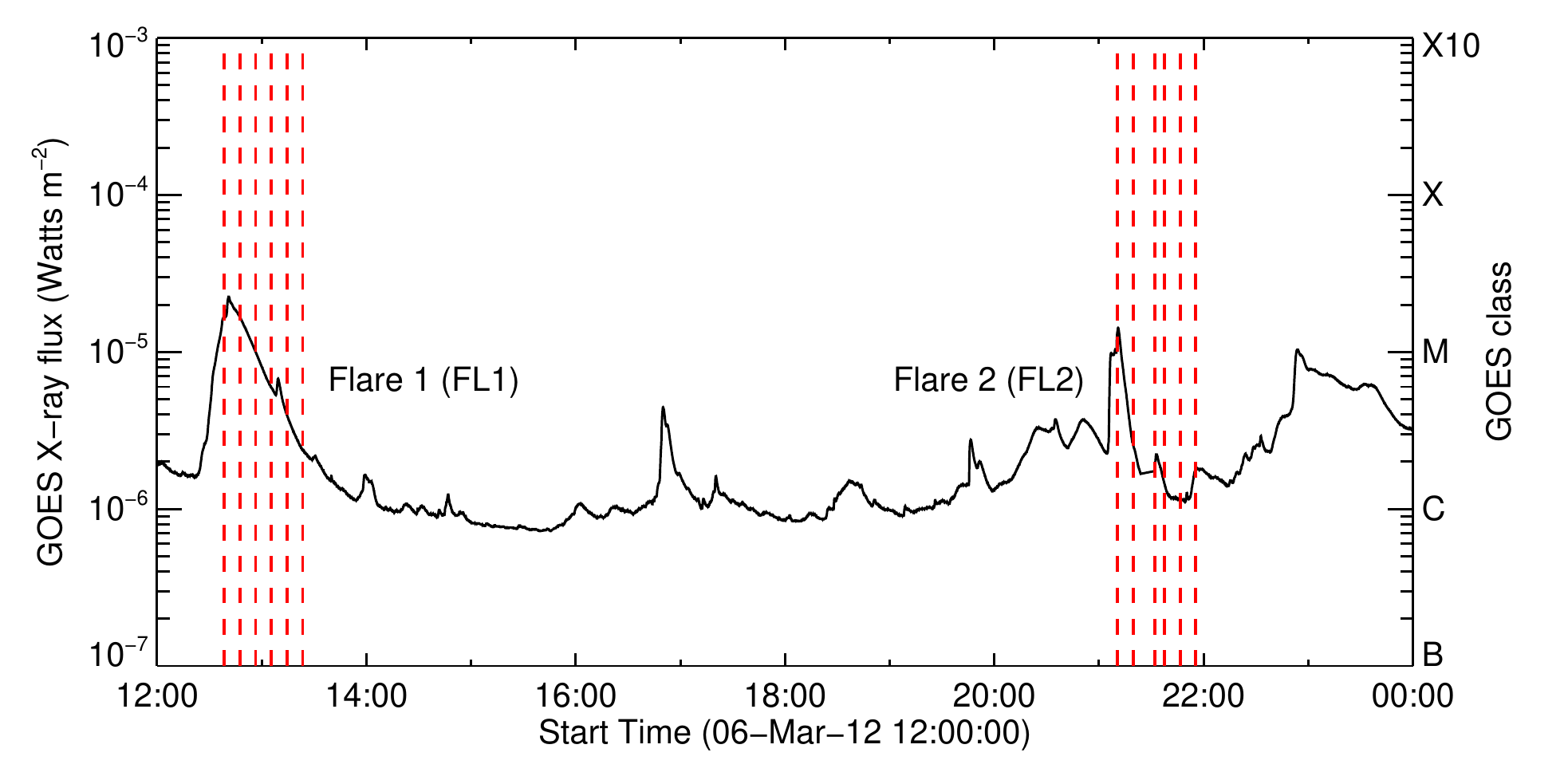}
\includegraphics[scale = 0.83]
{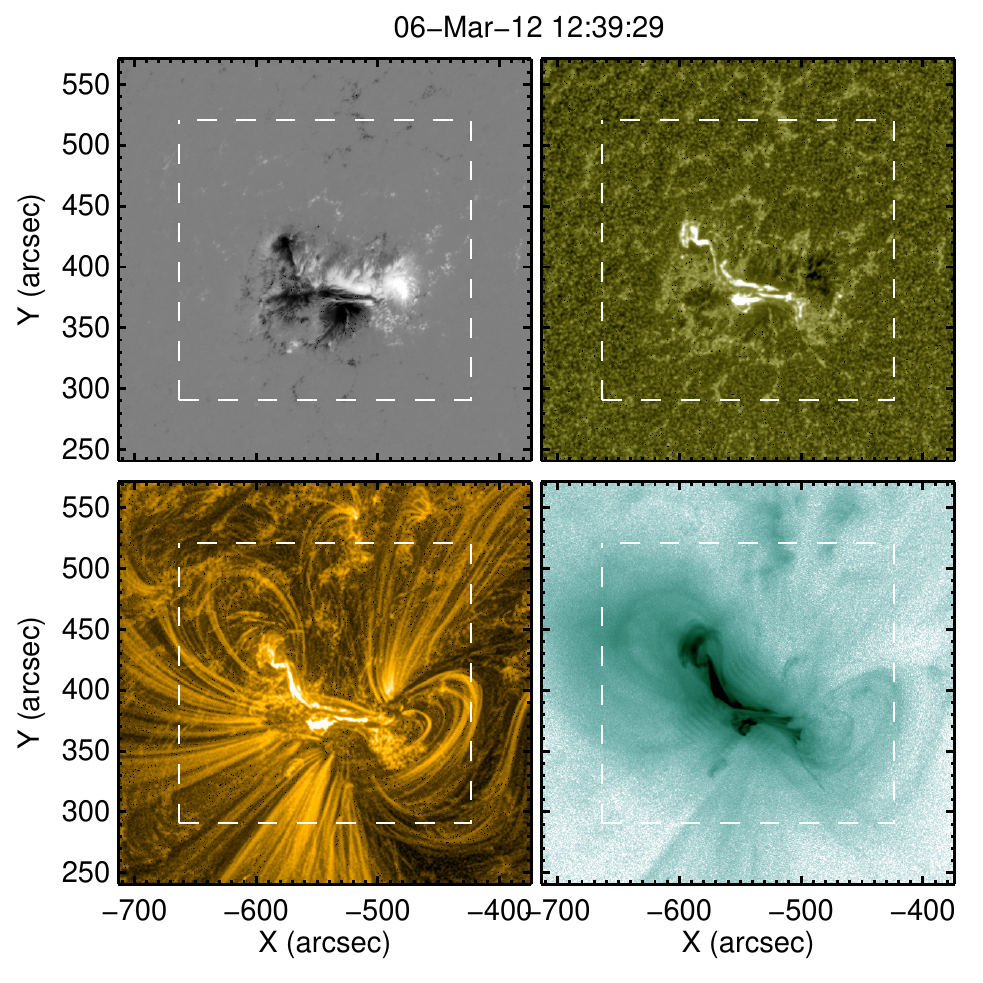}
\caption{Top panel:  \emph{GOES} soft X-ray curve from 12:00 UT on 2012 March 6 to 00:00 on March 7 with \emph{Hinode}/EIS raster times indicated by the dashed red lines.  Flare 1 (M2.2) peaked at $\sim$12:37 UT and Flare 2 (M1.4) peaked at $\sim$21.12 UT.
Bottom panel:  Clockwise from upper left: \emph{SDO}/HMI line-of-sight (LOS) magnetogram, \emph{SDO}/AIA 1600 $\angstrom$, 94 $\angstrom$, and 171 $\angstrom$ maps of AR 11429 at 12:39 on 2012 March 6 overplotted with \emph{Hinode}/EIS field of view (dashed white box).  An animation of the \emph{SDO} images in the bottom panel is available (Movie$\_$1.mp4).  Note that the cadence of the movie is 45 sec for the flare periods but is five min from 14:00 UT to 20:15 UT. \label{fig:multi}}
\end{figure}

The pre-flare coronal configuration of AR 11429 was a highly sheared structure.
The hot-channel passbands e.g., 94 $\angstrom$ and 131 $\angstrom$ of the \emph{Solar Dynamics Observatory's} Atmospheric Imaging Assembly (\emph{SDO}/AIA; \cite{lemen12}) have sheared bright loops evident on March 5.
One day later, the bright sheared loop system has evolved into a highly twisted `corkscrew' on its NE end at the beginning of the rise phase of flare FL1. 
Figure \ref{fig:multi} (bottom panel) shows a frame from the included movie Movie$\_$1.mp4 at 12:39 UT on March 6 in which the `corkscrew' feature is prominent in each of the \emph{SDO}/AIA passbands (1600 $\angstrom$, 171 $\angstrom$, and 94 $\angstrom$). 
Movie$\_$1.mp4 spans from 11:44 UT to 22:29 UT, covering both confined flares at a cadence of 45 sec but  the cadence is 5 min from 14:00 UT to 20:15 UT.

\begin{figure}[t!]
\centering
\includegraphics[scale = 0.25]{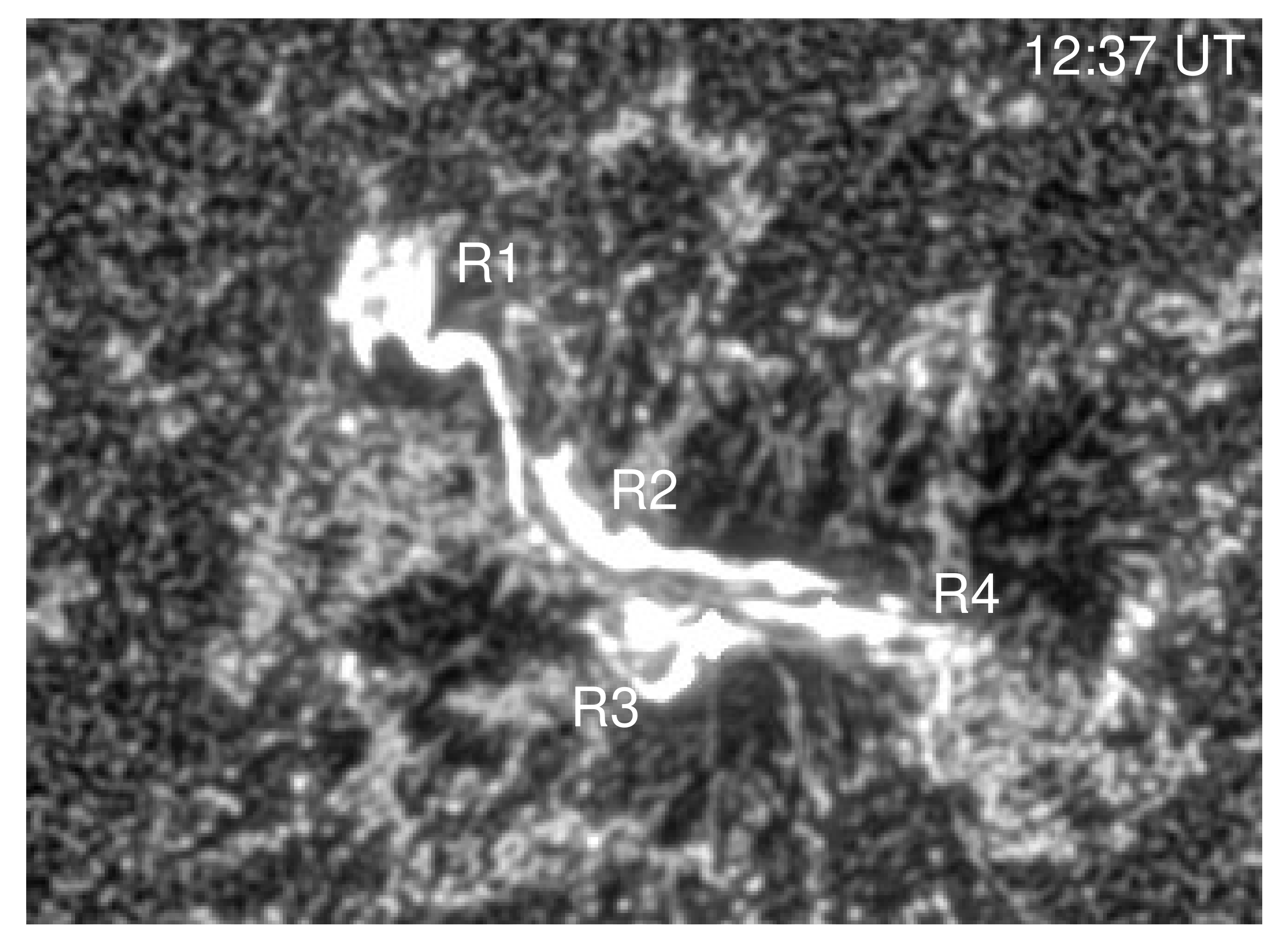}
\caption{Location of ribbon pairs R1--R2 and R3--R4 in \emph{SDO}/AIA 1600 $\angstrom$ image at 12:37 UT during flare FL1.  \label{fig:ribbons}}
\end{figure}

\cite{syntelis16} conducted a spectroscopic analysis of the pre-eruptive configuration of AR 11429 prior to the eruption of two CMEs that occurred early on 2012 March 7.
Using the same \emph{Hinode}/EIS data as in our present study, they found substantial spectroscopic evidence for the presence of a hot flux rope in the NE section of the AR which formed during flare FL1.
Two distinct plasma components were identified by \cite{syntelis16}, one at 1.6--2.5 MK (log$_{10}$ \emph{T} = 6.2--6.4 K) and the other at 6.3--12.6 MK (log$_{10}$ \emph{T} = 6.8--7.1 K).
The hotter component contained regions of enhanced non-thermal line broadening, relatively strong Doppler upflow velocities, and lower plasma densities. 
Collectively, these spectral parameters are characteristic of hot flux ropes.

\subsection{Flare Ribbons}
The lower solar atmosphere evolved in parallel with the corona during confined flare FL1. 
Elongated bright ribbons appear early in the rise phase of FL1, however, there is evidence of pre-flare heating at these locations in the \emph{SDO}/AIA 1600 $\angstrom$ and 171 $\angstrom$ passbands (see included movie Movie$\_$1.mp4).
At the peak of the flare, two pairs of flare ribbons, R1--R2 and R3--R4, are visible in the \emph{SDO}/AIA 1600 $\angstrom$ passband in Figure \ref{fig:ribbons}.
Ribbon pair R1--R2 is centered on the N--S aligned, near-vertical section of the polarity inversion line (PIL; cf. \emph{SDO's} Helioseismic and Magnetic Imager \citep[HMI;][]{scherrer12} magnetogram in Figure \ref{fig:multi}) in the NE and the second pair, R3--R4, runs along the E--W aligned, horizontal segment of the PIL in the SW of the AR.
Ribbon pair R1--R2 appears $\sim$1 min before R3--R4.
They are brightest at the flare peak (12:37 UT in Figure \ref{fig:ribbons}) and gradually decrease in intensity during the decay phase.

There was continuous, low-level reconfiguration of the corona in the period between the two flares when the main activity shifted to the vicinity of the ribbon pair R3--R4 of flare FL1, along the horizontal portion of the PIL.  
At 20:43 UT, brightenings begin to reappear in \emph{SDO}/AIA 1600 $\angstrom$ passband in the NE section of the AR while the main activity continues in the SW (see included movie Movie$\_$1.mp4).
The ribbons have faded in the NE by 21:06 UT, at the same time that FL2 begins in the vicinity of the former ribbons R3--R4.
We observe highly packed ribbons over the magnetically complex horizontal PIL during the rise phase.
Activity continues in the central part of the AR throughout the decay phase of FL2.

\subsection{Magnetic Field Evolution}\label{sec:Bfield}

\begin{figure*}[htp]
\centering
\includegraphics[width=0.75\textwidth]{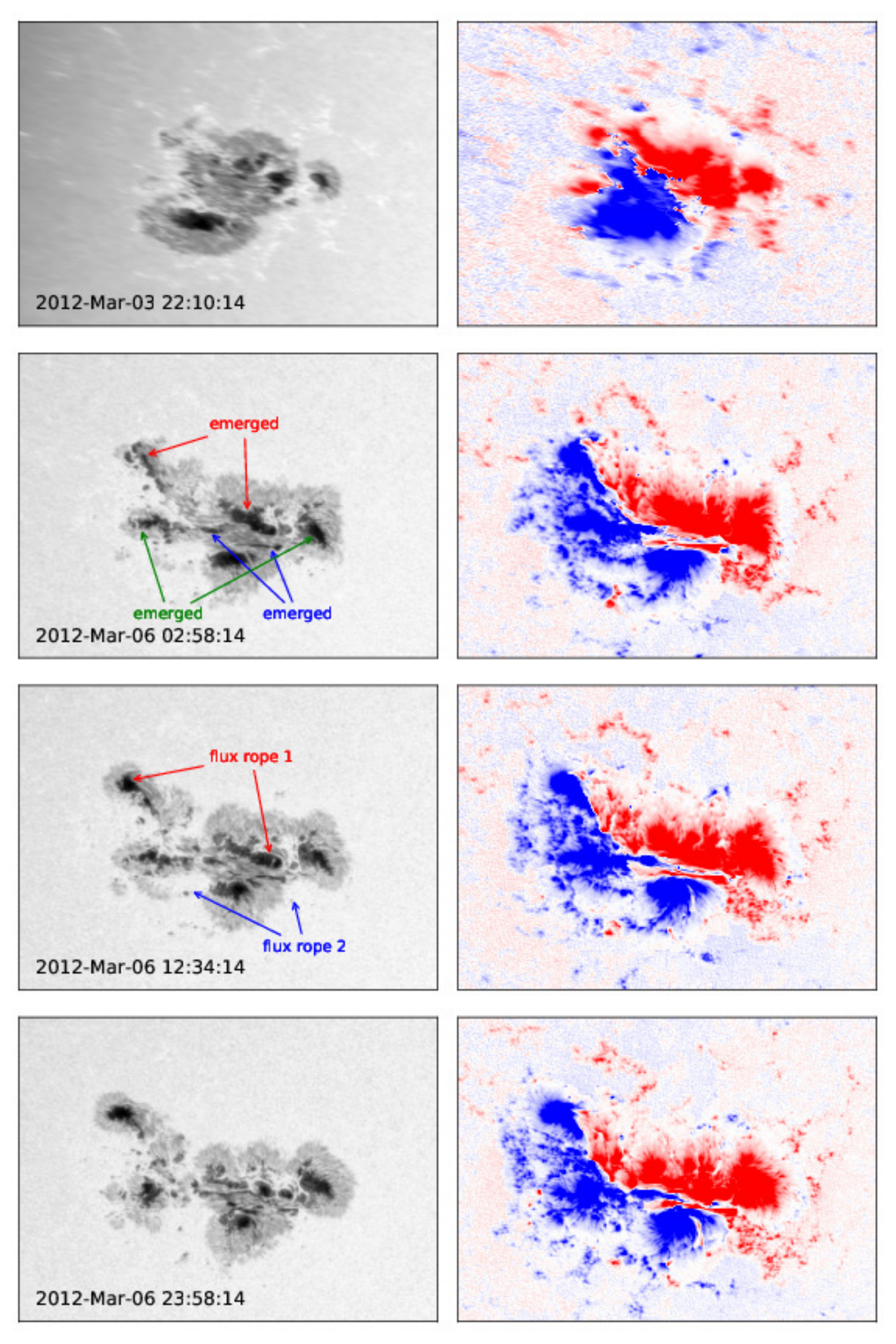}
\caption{Maps of \emph{SDO}/HMI SHARP CEA continuum (left panel) and radial magnetic field, $\mathbf{B}_{r}$ (right panel).  The positive/negative red/blue radial magnetic field is saturated at $\pm$1500 G.  Key flux emergence episodes are labeled using red, blue and green text (see continuum map at 02:58 on March 6). Footpoints of magnetic flux ropes 1 $\&$ 2 (red and blue, respectively) in the continuum map at 12:34 on March 6 correspond to the NLFF extrapolation in Figure 11 of \cite{chintzoglou15}.  MFR 1 is rooted in the sunspot umbrae in the north of AR 11429 whereas the footpoints of MFR2 are located further south in the quiet Sun. Included Movie$\_$2.mp4 is an animation of this figure.  In the movie, the locations of IFIP-biased and photospheric plasma observed by \emph{Hinode}/EIS are identified with arrows.  \label{fig:cont_bfield}}
\end{figure*}

When AR 11429 rotated onto the disk on 2012 March 3, it was already a mature sunspot group containing several umbrae in a common penumbra. 
The magnetic structure of the AR was bipolar, with an anti-Hale orientation for the N hemisphere (Figure \ref{fig:cont_bfield}, continuum image at 22:10 UT on March 3). 
The orientation of the bipolar fields also deviated significantly from Joy's law \citep{hale19}. 
Over the next few days, major flux emergence took place along the AR's inversion line, with the opposite-polarity new flux concentrations diverging at roughly a right angle to the line connecting the pre-existing spots. 
By March 6, the magnetic structure became $\alpha\beta\gamma$, with two--three major bipoles still in emergence. 

In Figure \ref{fig:cont_bfield}, 2$^{nd}$ row from the top, the main emerging bipoles are indicated with arrows of different colors. 
Since the emerging flux was highly sheared, the magnetic inversion line maintained its NE--SW orientation from March 3 throughout the emergence process. 
High shear in the emerging fields was evidenced in the radial magnetic field as a \emph{yin-yang} magnetic polarity pattern \citep[magnetic tongues;][]{luoni11}, indicating strong negative, left-handed helicity in the emerging flux.
Early on March 6, \cite{chintzoglou15}, using SDO/HMI SHARP vector magnetograms, found the mean shear angle along the PIL to be around 68$\degree$, which showed a slow increase throughout the day. 
\cite{chintzoglou15} pointed out that there were two sources of shear: highly sheared flux emergence in the NE part of the AR, and shearing motions between emerging fields (their positive polarity) and the pre-existing negative-polarity spot in the middle of the AR, where the magnetic PIL was nearly of E--W orientation, and had a quadrupolar structure (cf. Figure \ref{fig:cont_bfield}, right column). 

\cite{takasao15} carried out magneto-hydrodynamic (MHD) simulations showing that the emergence of a subsurface-kinked flux rope can spontaneously form a quadrupolar non-Hale, non-Joy AR with multiple magnetic inversion lines in its midst. 
The complex inversion line structure forms in their simulation by the submergence of emerged fields. 
This simulation gives a good description of AR 11429 and its multiple inversion lines on March 6 (cf. Figure \ref{fig:cont_bfield}, right column).

Using non-linear-force-free (NLFF) magnetic extrapolations, \cite{chintzoglou15} found two flux ropes on March 6 in AR 11429: magnetic flux rope 1 (MFR1) in the NE emerging fields, while flux rope 2 (MFR2) formed around the pre-existing negative spot, where strong shearing was observed around the complex E--W inversion line area. 
MFR1 had its footpoints in the forming new umbrae, while the latter (MFR2) flux rope had its footpoints over quiet-sun areas. 
The locations of the flux rope footpoints are indicated in the continuum image at 12:34 UT on March 6 in Figure \ref{fig:cont_bfield}.
\cite{chintzoglou15} suggested that MFR1, and probably MFR2, formed during the M2.2 flare (FL1) that we analyze. 
The flare ribbon pair R1--R2 is linked to the formation of MFR1, while ribbon pair R3--R4 to MFR2 (cf. Figure \ref{fig:ribbons} and Section 2.3).  

For about a day prior to FL1, the highly sheared divergence of opposite polarities emerging along the PIL led to an inflow of negative magnetic flux moving toward the isolated pre-existing spot at the eastern footpoint of MFR1 (R1).
Flux approached and collided with the spot, forcing the coalescence of the smaller flux fragments into a growing, strongly coherent umbra surrounded by a common penumbra.
At the western footpoint (R2), the emerging positive flux fragments crashed into and coalesced with the main positive polarity in the center of the active region.
Repeated flux emergence episodes drove umbral coalescence until $\sim$13:00 UT during FL1.
By $\sim$15:00 UT, the umbra at R1 had ceased coalescing and by $\sim$16:00 UT, the umbra at R2 started to break apart/decay.
This process is evident in Movie$\_$2.mp4 beginning at 11:22 UT on March 5 where arrows indicate each footpoint region in the continuum and radial field images. 

\bigskip
\section{\emph{Hinode}/EIS Observations of AR 11429} \label{sec:obs}

The observations featured in Figures \ref{eis_panel_early} and \ref{eis_panel_late} were acquired while \emph{Hinode}/EIS was operating in an autonomous observing mode during a Major Flare Watch campaign.  
When an intense brightening was detected in the He {\sc ii} 256.32 $\angstrom$ lines, a high cadence flare response study was triggered.
A series of 6 rasters at a cadence of nine minutes was 
run for each flare.
Observations spanned 12:38 UT to 13:32 UT (FL1) and from 21:10 UT to 22:04 UT (FL2) on 2012 March 6.
EIS captured the peak and the decay phase for both flares (see top panel of Figure \ref{fig:multi}).
Some of the key details of the flare response study are listed in Table \ref{tab:study}.

\begin{table}
	\centering
	\caption{\emph{Hinode}/EIS Study Details.}
	\label{tab:study}
	\begin{tabular}{ll} 
		\hline
		Study name & FlareResponse01\\
		IFIP Emission Lines & Ca {\sc xiv} 193.87 $\angstrom$\\
        &Ar {\sc xiv} 194.40 $\angstrom$\\
       	Field of view & 240$\arcsec$ $\times$ 304$\arcsec$\\
		Rastering & 2$\arcsec$ slit, 80 positions, 3$\arcsec$ coarse steps \\
        Exposure Time &5 s\\
		\hline
	\end{tabular}
\end{table}
\bigskip
\subsection{Ar {\sc xiv}/Ca {\sc xiv} Intensity Ratio}\label{sec:ratio}
The EIS flare response study contains two lines that are suitable for measuring plasma composition at temperatures higher than those expected in non-flaring ARs: low-FIP Ca {\sc xiv} (FIP = 6.11 eV) at 193.87 $\angstrom$ and high-FIP Ar {\sc xiv} (FIP = 15.76) at 194.40 $\angstrom$ \citep{feldman09}.
Both lines are relatively strong, close in wavelength, and contain no strong blends.
The two ions are formed in ionization equilibrium at similar temperatures of $\sim$3.5 MK (log$_{10}$ \emph{T} = 6.55 K) \citep{feldman09,doschek15}, and they
have very similar emissivity temperature dependences: the theoretical coronal ratio is about 0.25$\pm$0.10 over the temperature range 1.6--6.3 MK (6.2 $\leq$ log$_{10}$ \emph{T} $\leq$ 6.8 K) in coronal conditions at an electron density of log$_{10}$ \emph{N} = 10.
However, the ratio rises to levels exceeding 0.50 at a temperatures above $\sim$8 MK (log$_{10}$ \emph{T} = 6.9 K). 
(See Figure \ref{fig:goft} of the Appendix for a plot of the ratio of Ar {\sc xiv} and Ca {\sc xiv} contribution functions at different densities calculated using the abundances given below).
With similar contribution functions, the intensity ratio of these lines can be used to determine (I)FIP bias levels.

In line with \cite{doschek15,doschek16}, we use Log$_{10}$ abundance values, relative to H (with Log$_{10}$ (H density) set to 12), as follows:  corona--Ca = 6.93 \citep{feldman92} and Ar = 6.50 and photosphere--Ca = 6.33 \citep{caffau11} and Ar = 6.50 \citep{lodders08}, yielding a coronal FIP bias of $10^{0.6}$ = 4.
The CHIANTI Atomic Database, Version 8.0 \citep{dere97,delzanna15} was employed to carry out the calculations of the contribution functions using these abundances.
The abundances of Ar and Ca affect mostly the magnitude of the respective contribution functions but not their shapes as a function of temperature, so the intensity ratio allows us to determine the relative abundances of Ar to Ca.
Typically, the FIP bias is the ratio of the low-FIP element's coronal to photospheric abundances and the high-FIP element's coronal to photospheric abundances, however, the convention for IFIP bias is to invert the ratio.
We adopt the latter convention such that high-FIP Ar {\sc xiv}/low-FIP Ca {\sc xiv} line intensity ratio $>$ 1 indicates the IFIP effect, = 1 is unfractionated photospheric plasma, and $<$ 1 is the solar FIP effect.
The estimated uncertainty based on an intensity error of 20$\%$ is $\pm$0.28.

All Hinode/EIS data were reduced using the eis$\_$prep routine available in Solar SoftWare \citep{freeland98}.
The routine converts the CCD signal in each pixel into calibrated intensity units and removes/flags cosmic rays, dark current, dusty, warm, and hot pixels.
The original \emph{Hinode}/EIS calibration was used instead of the newer calibrations of \cite{delzanna13a} and \cite{warren14b} because the lines are close in wavelength and within the same spectral window. 
The ratio of the corrected intensities using either calibration is within 2$\%$ of the original.
We fit three Gaussians to the Ar {\sc xiv} 194.40 $\angstrom$ line to remove two unidentified weak lines in its blue wing \citep{brown08,doschek15} and to the spectral region around the Ca {\sc xiv} 193.87 $\angstrom$ line to separate it from two nearby lines.
Two sample spectra for the spectral window are shown in Figure \ref{fig:wavelength} of the Appendix, one of the FIP effect and the other of the IFIP effect.
Typically, the unidentified weak lines in the blue wing of the Ar {\sc xiv} are evident in the FIP effect spectra but are negligible in the IFIP effect spectra.
Columns 3 and 4 of Figures \ref{eis_panel_early} and \ref{eis_panel_late} show the Ar {\sc xiv}/Ca {\sc xiv} intensity ratio maps.

In this work, we have followed the methodology of \cite{doschek15,doschek16,doschek17} in producing the Ar {\sc xiv}/Ca {\sc xiv} intensity ratio maps.
They give a full account of the assumptions and issues concerning this EIS composition diagnostic.
We refer the reader to their extensive discussions contained in this series of papers, especially \cite{doschek17}.  
This method has the advantage of being simple, but does not fully account for the (relatively weak) temperature and density sensitivity of the
ratio. 
To alleviate any concerns that these effects might explain the detection of patches of inverse FIP, as an independent check, we have analyzed one of the IFIP patches using the more complete method of computing the FIP bias from a differential emission measure (DEM) analysis. 
The DEM was inferred by measuring the electron density and using it to compute contribution functions for a set of emission lines covering a wide range of temperatures (Fe {\sc viii-xxii}, Ca {\sc xiv--xv}). 
The ratio of the calculated to observed Ar {\sc xiv} intensity gives the FIP bias. 
This analysis showed that accounting for the temperature and density sensitivity increased the magnitude of the inverse FIP effect. 
The reason is that the Ar {\sc xiv} line is brighter than expected from theory, so taking the observed Ar {\sc xiv}/Ca {\sc xiv} ratio underestimates the inverse FIP effect. 
In addition, when the contribution functions are convolved with the DEM, it appears that the two lines are formed closer to 4.5 MK (log$_{10}$ \emph{T} = 6.65 K), which falls within the temperature range of the ratio values quoted above. 
They are also formed over a much narrower temperature range than expected from theory, so any larger variations in the ratio outside of the temperature range 3.2--6.3 MK (6.5 $\leq$ log$_{10}$ \emph{T} $\leq$ 6.8 K) are not important in this event (see plots of the ratio in Figure \ref{fig:goft} of the Appendix).
The theoretical Ar {\sc xiv} and Ca {\sc xiv} contribution functions (G(T)) along with their contribution functions convolved with the DEM are shown in Figure \ref{fig:norm_goft} of the Appendix.

\subsection{IFIP Plasma Evolution During Flare FL1}
Figure \ref{eis_panel_early} is composed of \emph{Hinode}/EIS and \emph{SDO}/AIA images during flare FL1 as follows: columns 1 and 2 contain EIS Ar {\sc xiv} and Ca {\sc xiv} intensity images; the Ar {\sc xiv}/Ca {\sc xiv} ratio maps are in columns 3 and 4, without and with \emph{SDO}/HMI line-of-sight (LOS) contours of $\pm$500 G, and \emph{SDO} 1600 $\angstrom$ and 94 $\angstrom$ intensity images are displayed in the last two columns.
The saturation levels are fixed for each column of intensity images to discern the relative changes in brightness with time for a given wavelength or passband.

The first EIS observation at 12:38 UT coincides with the flare peak. 
At that time, the AR contains plasma that is predominantly of coronal composition.
FIP bias ranges from 0.25--0.40 (using the ratio of high-FIP element Ar {\sc xiv} over low-FIP element Ca {\sc xiv} but equivalent to 2.5--4 in the solar FIP bias convention as stated in the previous section).
The highest FIP bias ($\sim$0.25) is located in the N--S directed bright loops evident in the EIS Ar {\sc xiv}, Ca {\sc xiv}, and AIA 94 $\angstrom$ intensity images.
Nine minutes later at 12:47 UT, the extent of the strong FIP bias material has spread as the bright loops expand during the confined eruption.
This is also when the first highly localized patches of IFIP plasma composition are visible in the EIS Ar {\sc xiv}/Ca {\sc xiv} ratio images.
Both IFIP regions are bordered by a `moat' of photospheric composition which in turn is surrounded by coronal plasma composition. 
Throughout the remainder of the decay phase of FL1, the western IFIP patch is still clearly visible from 12:56 UT to 13:23 UT, however, the eastern patch evolves from IFIP to photospheric composition (cf. Ar {\sc xiv}/Ca {\sc xiv} intensity ratio images at 13:14 UT and 13:23 UT in column 3 of Figure \ref{eis_panel_early}).
IFIP values are in the range [1.5, 2.1].
The IFIP plasma appears within the 9-minute time cadence of the EIS rasters and then rapidly evolves toward photospheric composition within 18 minutes during the decay phase of the M2.2 flare.

Figure \ref{eis_panel_early}, column 4, shows that the IFIP patches are located at the footpoints of the bright loops connecting the strong magnetic field concentrations of the northern bipole.
We have overlaid green contours of the strong IFIP regions observed at 12:47 UT on the \emph{SDO}/HMI continuum image at the corresponding time in Figure \ref{fig:contours} to emphasize the very specific position of the anomalous composition within the strong field.
The IFIP plasma is located entirely within the umbrae of the northern bipole.
Furthermore, the flare ribbons of FL1 also cross/end at the same places as the IFIP patches over the umbrae (cf.  the intersection of the orange contours of the flare ribbons and the green contours at the umbrae associated with ribbon R1 in the NE and R2 along the horizontal PIL).
In summary, IFIP composition is found over umbral areas crossed by flare ribbons.
The footpoints of flux rope MFR1 of \cite{chintzoglou15} are rooted in the same umbrae of the emerging bipole where we observe the IFIP plasma.

\subsection{Photospheric Plasma Composition of Flare FL2}
\begin{figure*}[htp]
\centering
\includegraphics[scale = 0.45]{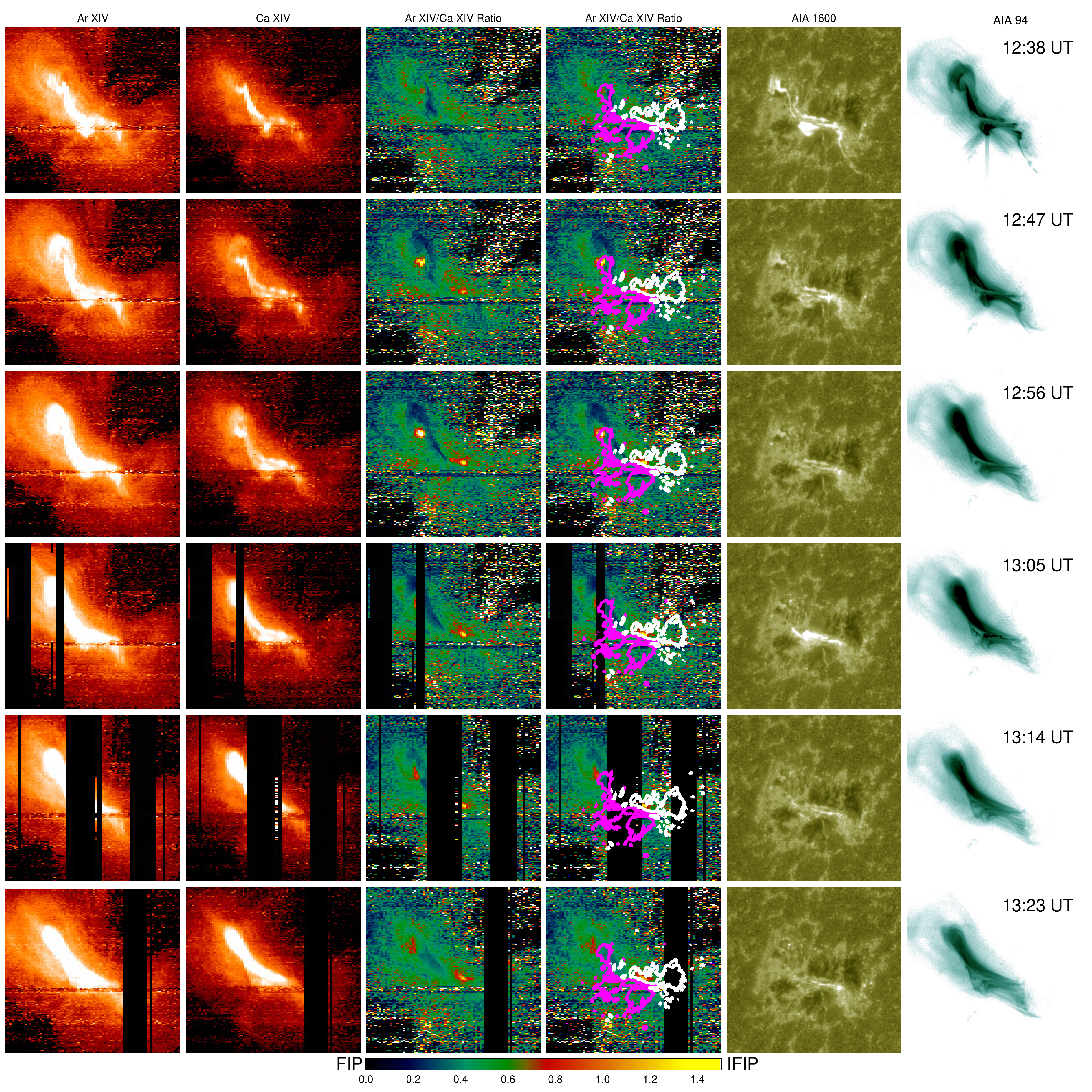}
\caption{Left to right:  \emph{Hinode}/EIS Ar {\sc xiv} 194.4 $\angstrom$~and Ca {\sc xiv} 193.87 $\angstrom$~intensity maps, Ar {\sc xiv}/Ca {\sc xiv} ratio maps without and with \emph{SDO}/HMI contours of $\pm$500 (white/purple), \emph{SDO}/AIA 1600 $\angstrom$~and 94 $\angstrom$~maps. (\emph{SDO}/AIA 94 $\angstrom$~map shown using reverse color table).  Top to bottom:  Observations are from 12:38 UT to 13:32 UT during flare FL1.  The color bar scale shows the FIP effect as blue/green, photospheric composition as orange, IFIP effect as yellow.  All \emph{Hinode}/EIS maps are co-aligned to \emph{SDO}/AIA and HMI maps at the times shown.  Black vertical stripes indicate data drop out periods.}\label{eis_panel_early}
\end{figure*}

\begin{figure*}[htp]
\includegraphics[scale = 0.45]{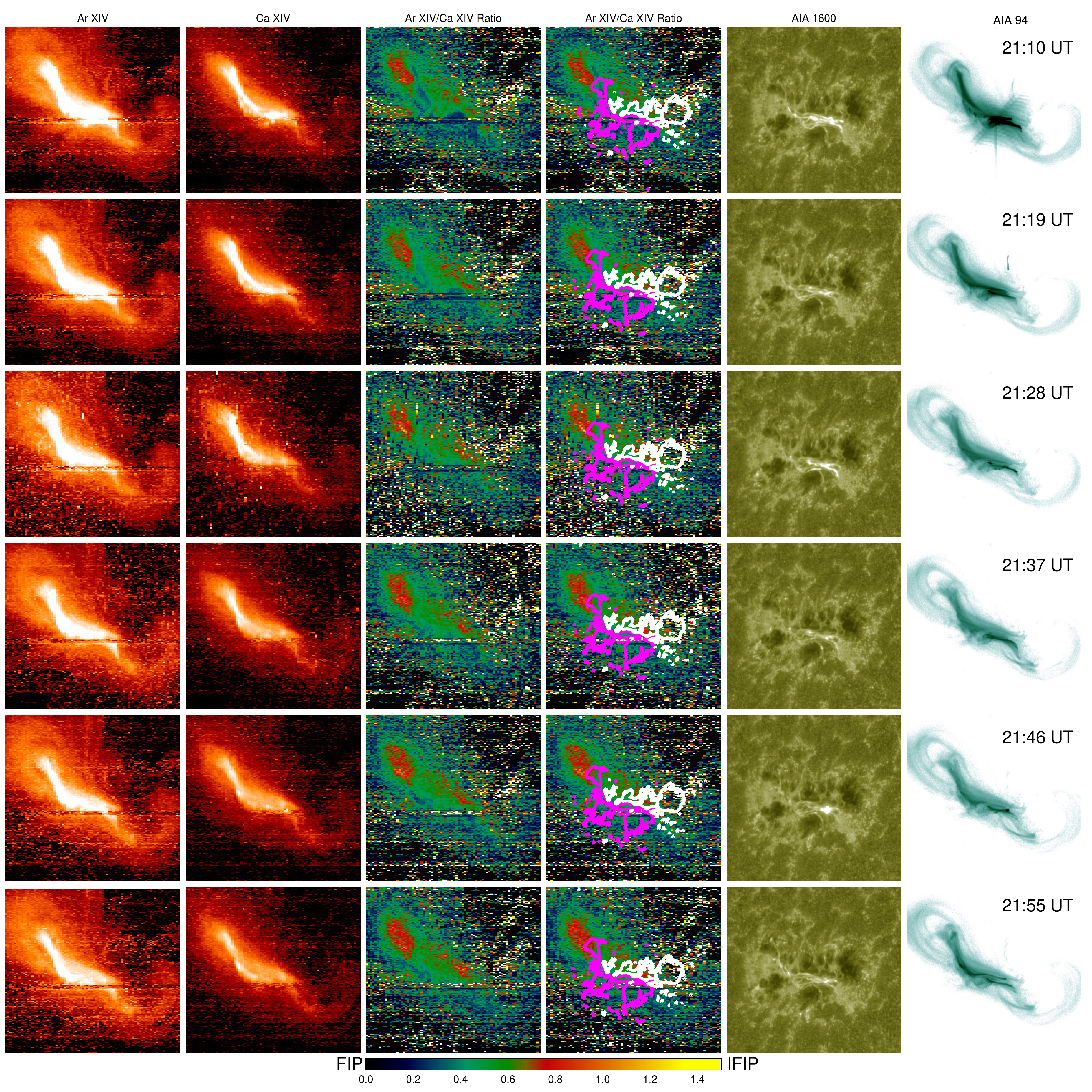}
\caption{Left to right:  \emph{Hinode}/EIS Ar {\sc xiv} 194.4 $\angstrom$~and Ca {\sc xiv} 193.87 $\angstrom$~intensity maps, Ar {\sc xiv}/Ca {\sc xiv} ratio maps without and with \emph{SDO}/HMI contours of $\pm$500 (white/purple), \emph{SDO}/AIA 1600 $\angstrom$~and 94 $\angstrom$~maps. (\emph{SDO}/AIA 94 $\angstrom$~map shown using reverse color table).  Top to bottom:  Observations are from 21:10 UT to 22:04 UT during flare FL2.  The color bar scale shows the FIP effect as blue/green, photospheric composition as orange, IFIP effect as yellow.  All \emph{Hinode}/EIS maps are co-aligned to \emph{SDO}/AIA and HMI maps at the times shown.}\label{eis_panel_late}
\end{figure*}

\begin{figure}[htp]
\centering
\includegraphics[scale = 0.27]{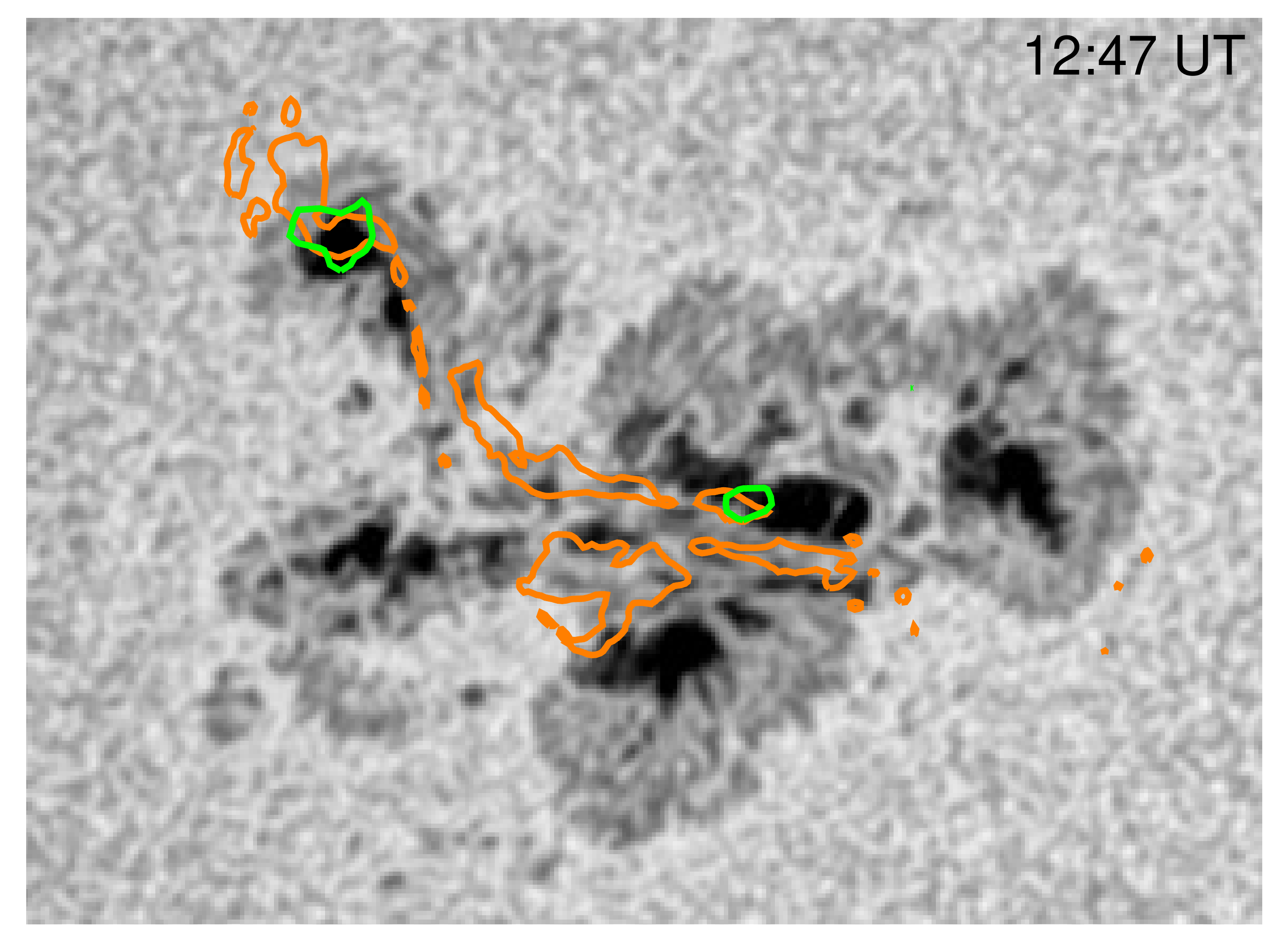}
\caption{\emph{SDO}/HMI Continuum at 12:47 UT overlaid with contours of IFIP (green) and flare ribbons (orange). The contours overlap at the umbrae associated with ribbon pairs R1--R2.  IFIP contours are from the \emph{Hinode}/EIS Ar {\sc xiv}/Ca {\sc xiv} intensity ratio map at 12:47 UT in Figure \ref{eis_panel_early}.  Flare ribbon contours are from the \emph{SDO}/AIA 1600 image $\angstrom$ at 12:37 UT in Figure \ref{fig:ribbons} adjusted for solar rotation.}\label{fig:contours}
\end{figure}

Figure \ref{eis_panel_late} contains the same series of images for flare FL2 as is used for FL1 in Figure \ref{eis_panel_early}.
Once again, the EIS observations began after the flare trigger was initiated and continued throughout and beyond the short decay phase of the flare.
The locations of the IFIP plasma within AR 11429 remained fixed throughout the decay phase of FL1 and appear to correspond to where photospheric-like composition is observed $\sim$8 hours later during flare FL2 (see the orange patches in the EIS Ar {\sc xiv}/Ca {\sc xiv} intensity ratio maps in Figure \ref{eis_panel_late}).
Very little composition evolution is evident during FL2. 
The anomalous composition is no longer found to be in distinct IFIP patches; rather it has near-photospheric composition and is more dispersed especially in the eastern region.
In the western region, much of the plasma over the positive polarity has evolved back toward coronal composition.  
The umbra where the eastern footpoint of MFR1 is rooted remains a coherent structure during and after FL2, though it has stopped coalescing, whereas, the umbra at the other (western) footpoint has broken up and dispersed (cf. the continuum images at 12:34 and 23:58 of Figure \ref{fig:cont_bfield} and see the location indicated by the right arrow in each arrow-pair of Movie$\_2$.mp4). 

\section{Discussion and Interpretation} \label{sec:disc}

\subsection{Summary of Plasma Composition Evolution in the Confined Flares}
In this study, we have analyzed the evolution of plasma composition in the highly active AR 11429 observed by \emph{Hinode}/EIS during two confined M-class flares.
During the first confined flare, FIP bias plasma comprises the AR when the flare is at peak intensity.
Minutes into the flare's decay phase, IFIP bias plasma appears at the footpoints of bright flare loops within the AR.
The IFIP patches are surrounded by a ring of photospheric material.
Co-temporally, the spatial extent of FIP bias increases in the bright flare loops. 
By the time the \emph{GOES} soft X-ray intensity has returned to pre-flare levels, there is asymmetric composition evolution in the flare loop footpoints; the plasma at the eastern footpoints evolves to photospheric composition whereas at the western footpoints, it maintains IFIP composition encircled by photospheric plasma composition.
\emph{Hinode}/EIS observes these extremes in plasma evolution in less than one hour, primarily within the decay phase of the flare.
Approximately seven hours later the second confined flare occurs and photospheric plasma composition is present in the vicinity of the flare loop footpoints where IFIP plasma was observed during the first flare.

Distinct IFIP patches occurred near the footpoints of one of the two magnetic flux ropes identified by \cite{chintzoglou15} where flare ribbons cross the umbrae of the emerging bipole in the northern section of the AR.
These are very particular locations within the unusually complex magnetic configuration of the AR.
The fact that the IFIP plasma is only observed for a short time during the decay phase of a moderate flare in highly localized places within the AR's magnetic field raises key questions as to what roles, if any, the AR's magnetic field configuration and flaring activity have in the creation and observation of anomalous plasma composition.
\bigskip
\bigskip
\subsection{What is the significance of the emergence of highly sheared magnetic field and coalescing sunspots?}
Such field represents different strands of highly sheared field that are converging towards each other to form sunspots, and therefore meet below the photosphere/chromosphere in the location of the coalescing umbrae. 
The high shear in coalescing strands of the same magnetic polarity suggests that the strands with a non-zero component of anti-parallel magnetic field have the possibility for sub-chromospheric magnetic reconnection when brought together.
In the high-$\beta$ plasma regime of the photosphere/low chromosphere, upward-moving acoustic waves are generated which can mode convert to fast mode waves at $\sim$1 Mm above the photosphere in the chromosphere where plasma-$\beta$ is equal to unity \citep{bourdin17}. 
Over regions of high magnetic field concentration such as sunspots, the transition to a low-$\beta$ plasma occurs at lower heights within the photosphere. 
The atmospheric models of \citet{avrett15}, constrained by observations, show a plasma pressure lower by a factor of 20 to 70 in a sunspot compared to the quiet Sun. 
This quiet Sun pressure model is close to the one used by \citet{gary01} to derive the plasma-$\beta$ versus height. 
With a field strength of 2500 G, \cite{gary01} found plasma-$\beta$ to be around or below 0.2 within the photosphere. 
With the  new sunspot atmospheric model of \citet{avrett15}, the plasma-$\beta$ is a factor between 20 to 70 lower, setting the sunspot umbra well in the low plasma-$\beta$ regime. 
Furthermore, the umbral magnetic field expands less with height than does the penumbral field. 
Consequently in the decreasing density, the Alfv\'en and fast mode speeds increase faster in the umbra, leading to the strongest refraction of fast mode waves, and the strongest downward ponderomotive acceleration.

This is significant in the context of the ponderomotive force fractionation model \citep{laming12,laming15,laming17}.  
When Alfv\'en wave flux originates in the corona, the ponderomotive force points upward, bringing low-FIP elements up from the chromosphere so that the FIP effect is observed.   
Fast mode waves coming from below the chromosphere means the ponderomotive force is directed downward thereby depleting low-FIP elements from chromospheric plasma.
This is consistent with \cite{brooks18b} who found that depletion of low-FIP elements instead of enhanced abundances of high-FIP elements yields IFIP plasma in post-flare loops.

In the Laming model, Ar/Ca is fractionated between just above the plasma-$\beta$ = 1 layer, and below a height of $\sim$1 Mm where H starts to be ionized. 
The plasma-$\beta$ = 1 layer with 300 G is in the photosphere. 
The deeper this layer lies, the more likely acoustic waves generated by sub-photospheric reconnection will mode convert to magnetoacoustic or fast mode waves when reflected/refracted at high density gradients while H is still neutral, and causes the IFIP fractionation. 
In the case of AR 11429, the umbral magnetic field exceeds 500 G, therefore it is plausible that the plasma-$\beta$ = 1 layer could be low down in the photosphere or even below that as discussed above, enabling the IFIP fractionation to take place.

Chromospheric dynamics generally occur on timescales much faster than those for ionization and recombination, and so the chromospheric ionization
balance is almost static, although elevated from that expected in equilibrium \citep{carlsson02}. 
This is reproduced in the \citet{avrett08,avrett15} models. 
A more important concern resulting from chromospheric dynamics would be the effect on the wave physics, especially in the low chromosphere
where the IFIP fractionation occurs.
We speculate that the extra wave interactions with density structures would increase the reflectivity of the chromosphere to Alfv\'en and fast mode waves, thus reinforcing our cconclusions. 
Extra dynamics in the chromosphere are probably required for IFIP, 
since it is worth noting that waves generated from solar or stellar convection are not strong enough, by an order of magnitude in amplitude, or two orders of magnitude in energy, to cause sufficient IFIP fractionation. 
Typical turbulent amplitudes are of order 1 km s$^{-1}$ \citep{bruntt10} and relatively constant with stellar spectral type \citep{kjeldsen11,chaplin09,baudin11}, while around 10 km s$^{-1}$ is necessary to produce sufficient ponderomotive acceleration \citep{laming15}.

The inference in this paper that subsurface reconnection generates the waves responsible for the IFIP fractionation is supported by surveys of FIP and
IFIP in stars. 
\cite{wood18} find FIP fractionation reducing and becoming IFIP fractionation in stellar coronae as the stellar spectral type becomes later (i.e., cooler).
Interestingly, the IFIP fractionation appears for stars where the magnetic field saturates when plotted against Rossby number, which is defined as the ratio of a star's rotational period to its convective turnover time \citep{reiners09}.
\cite{testa15} extend this plot to X-ray emission and FIP/IFIP fractionation against Rossby number. 
The saturation of magnetic field (or equivalently X-ray emission) implies the magnetic field generated by the rotation is being quenched, presumably by reconnection, which must presumably be subsurface reconnection because the X-ray emission also saturates. 
And, we emphasize, this subsurface reconnection coincides with the IFIP fractionation appearing in the coronae of these stars.

In the case of the pair of flare ribbons R3--R4 linked to flux rope MFR2 identified by \cite{chintzoglou15}, no anomalous plasma composition was observed at any time during either flare. 
MFR2 has footpoints in the quiet Sun \citep{chintzoglou15} and there were no coalescing umbrae observed in these locations (cf. the location of the footpoints of flux rope 2 in the continuum, radial magnetic field images at 12:34 UT on March 6 in Figure \ref{fig:cont_bfield} and the corresponding locations in Movie$\_$2.mp4).
Therefore, the conditions for sub-chromospheric reconnection and the resulting Alfv\'enic wave generation are not satisfied at least at the locations of the flare ribbons.
A comparison of the conditions associated with the two flux ropes highlights that the IFIP patches are found in very specific locations within the AR.

\subsection{Does flaring play a role in creating IFIP bias plasma?}

During a solar flare, magnetic reconnection high up in the corona accelerates electron beams and initiates Alfv\'en waves downward along magnetic loops connected to the chromosphere \citep[e.g.,][]{fletcher11,laming17,reep18}.
In the very dense chromosphere, kinetic energy is converted to thermal energy, causing the plasma to be heated up to temperatures of log$_{10}$ \emph{T} $\sim$ 7.0 K (10 MK). 
The overdense, heated plasma expands upward, and due to the high overpressure, fills the coronal loops driving mass flows from the chromosphere into the corona i.e., the process of chromospheric evaporation \citep{neupert68}. 

In most flares, evaporation leads to plasma of photospheric or near-photospheric composition as both low- and high-FIP elements are ionized and evaporated together (e.g., \cite{warren14a} and other references given in Section \ref{sec:intro}). 
The unfractionated plasma arises during flares because the upward flow speed is too fast for sufficient fractionation to occur \citep{laming09a}.
However, in FL1 we have chromospheric plasma that is depleted of low-FIP elements located above coalescing umbrae in the highly sheared emerging field.
The atomic masses of Argon and Calcium are similar, 39.948 amu and 40.078 amu, respectively, so that there is no preferential evaporation upflow of one element over another with the Ar {\sc xiv}/Ca {\sc xiv} intensity ratio composition diagnostic used in this study. 
Chromospheric evaporation brings up what is there, and in the very specific locations within AR 11429, this is low-FIP depleted plasma, consequently we observe IFIP patches in the corona.

The flare ribbons cover a variety of magnetic-field strengths from quiet-Sun to umbral (cf. the location of flare ribbon contours in Figure \ref{fig:ribbons} and radial magnetograms at approximately the same time in Movie$\_$2.mp4).
At those locations away from the strong field, the \emph{Hinode}/EIS observations show that the flare energy releases low-FIP-biased plasma first, creating strong low-FIP composition in flare loops as observed in FL1 and to a much lesser extent in FL2.
In Figure \ref{eis_panel_early}, the loops are filled with low-FIP enhanced plasma at 12:38 UT, at the time of the flare peak, and continue to be filled with the low-FIP biased plasma from 12:47 UT, when we first observe IFIP-biased plasma at the strong umbral field locations, until 13:05 UT. 

Over the coalescing umbrae, where the composition deeper in the chromosphere is depleted of low-FIP elements prior to the flare, the bottom of flare loops will be filled with plasma of IFIP composition. 
The IFIP patches appear to last as long as either chromospheric evaporation is triggered by ionizing electron beams produced by magnetic field reconnection in the corona and/or thermal conduction of coronal plasma persists \citep{bagala95,cheung18}.  
The only observations of high energy electrons of FL1 are from the \emph{Fermi Gamma-ray Space Telescope}.
The peak in the 25--50 keV energy bin occurs at the peak of the flare at 12:38 UT before returning to background levels by $\sim$12:45 UT (not shown here).
As \emph{Fermi} then entered its night phase, we do not have observations beyond 12:45 UT, however, the fact that the electron energy levels have returned to background levels by this time suggests that chromospheric evaporation is not driven by electron beams but rather by thermal conduction during the later phase of the flare.
In either case, once the process of chromospheric evaporation is finished, with radiative and conductive cooling taking place,  the IFIP plasma would no longer be observable using the Ar {\sc XIV}/Ca {\sc XIV} diagnostic ratio.
Therefore, the patches appear to decay within the gradual phase of the flare.

The EIS observations indicate that IFIP-composition plasma is present in the vicinity of footpoints in certain flare loops. 
The flaring reconnected loops inherit the composition of their progenitor loops, i.e., FIP composition. 
In addition, when the top of the chromosphere is being evaporated at the start of the flare, low-FIP-element enhanced plasma created during intermittent heating episodes enters the reconnected loops. 
Only later, when bremsstrahlung heating reaches deeper chromospheric layers, will the IFIP-biased plasma be evaporated.
As IFIP plasma is injected into the flare loops, chromospheric evaporation creates the strongest IFIP bias at loop footpoints.
At greater heights, plasma mixing creates a transition from IFIP bias through photospheric to FIP bias composition.
A composition gradient from photospheric to coronal (FIP effect) along flare loops was observed by \cite{doschek18} in an AR at the limb, but in this case there was no IFIP plasma at the loop footpoints.
In a quiescent newly emerged AR, \cite{baker13} also observed a composition gradient from the AR's loop footpoints to greater heights along the loops. 

Our scenario for the role of flaring activity in the creation and observation of IFIP plasma on the Sun is similar to the interpretation presented in \cite{laming09a} for a flare on the M dwarf star EV Lac which has an IFIP quiescent corona.
Element abundances during a moderate flare showed a near stellar photospheric composition.
\cite{laming09a} argue that the downward-directed ponderomotive force in the chromosphere of the IFIP-dominated star increases the heat conduction from the flare, consequently enhancing chromospheric evaporation and for active stars like EV Lac, stronger chromospheric evaporation leads to element abundance variation during flares.
However, in the case of the less active Sun, comparatively reduced evaporation leads to little variation in the abundances observed during flares, at least on large scales.

Furthermore, in a multi-wavelength study of a giant flare observed on active M dwarf CN Leo with an IFIP dominated corona, \cite{liefke10} found more than a twofold increase in the low-FIP Fe abundance during the flare i.e., the FIP effect, compared to the pre-flare quiescent abundance level.
The peak in Fe abundance enhancement approximately coincided with the flare's peak in soft X-ray emission.
By the end of the decay phase, the Fe abundance returned to the pre-flare level of the quiescent corona, similar to the progression of plasma composition we observed in the flare loops during FL1 in AR 11429.
In each case, the plasma appears to approach photospheric composition and then returns to pre-flare quiescent coronal composition during the time period of flare decay phase. 

\subsection{Why do we observe photospheric composition at the locations of the IFIP patches during the second flare?}

The Ar {\sc xiv}/Ca {\sc xiv} ratio maps in Figure \ref{eis_panel_late} contain photospheric plasma where IFIP patches had been observed during FL1.
As evidenced in Movie$\_$2.mp4 and described in Section \ref{sec:Bfield}, by the time of FL2 at $\sim$21:00 UT, the sunspot coalescence was completed and the umbrae had entered their decay phase, halting sub-chromospheric reconnection.  
Therefore, the fast mode wave flux arriving from below the chromosphere would have ceased, together with the resulting ponderomotive force, which had depleted low-FIP elements from these locations.
In effect, the IFIP mechanism was `turned off'.
Once the IFIP plasma is no longer generated, plasma mixing with the surrounding field containing low-FIP bias material would begin to change what we observe in the corona in these specific locations; from IFIP to photospheric composition in flaring loops.
Repeated flaring activity after FL1 may have provided a steady supply of unfractionated plasma to the vicinity of IFIP patches, accelerating the transition from IFIP to photospheric composition though this time period was not observed by \emph{Hinode}/EIS.  
The pattern of evolution is partly consistent with what \cite{nordon07,nordon08} observed on much larger scales in their study; flares tend to decrease both IFIP bias in the IFIP-dominated coronae of active stars and FIP bias in the coronae of solar-like stars \citep{testa10}.

\section{Conclusion}\label{sec:conclude}

In this case study, highly localized inverse-FIP composition patches are appearing during a confined flare in an overall FIP-bias dominated active region.  These patches evolve and fade during the decay phase of the flare.  
The IFIP patches are observed in highly sheared emerging flux over coalescing umbrae crossed by flare ribbons.
We propose that subsurface magnetic reconnection between coalescing umbrae led to an increased fast mode wave flux from below the fractionation height and resulted in the depletion of low-FIP elements.
When these coalescing umbrae with plasma composition depleted of low-FIP elements became footpoints of flare loops (i.e., they were crossed by flare ribbons), the chromospheric evaporation of the low-FIP depleted plasma led to the appearance of IFIP patches above these umbrae.
The IFIP patches were observed in the corona as long as chromospheric evaporation lasted. 
The insight gained from spatially resolved composition maps suggests that in Sun-as-a-star or stellar cases the composition of the chromosphere which is evaporated into flare loops may have local anomalies and a flare's effect on the overall coronal composition may depend on the filling factor of those anomalous locations. 
These findings are consistent with the Laming model as well as \cite{brooks18b}.  
In future work we will analyze the magnetic field evolution of the 8 active regions where IFIP plasma has been observed to determine how common are the characteristics seen in this case study when considering the entire sample.

\bigskip
\appendix
We address specific technical aspects of the Ar {\sc xiv}/Ca {\sc xiv} intensity ratio diagnostic described in Section \ref{sec:ratio}:  the effect of temperature on the ratio for temperatures exceeding log$_{10}$ \emph{T} = 6.8 K; the Gaussian fittings to the Ca {\sc xiv} spectral window for sample FIP and IFIP effect spectra; the contribution functions convolved with the differential emission measure (DEM).

\begin{figure}[H]
\centering
\includegraphics[scale = 0.45]{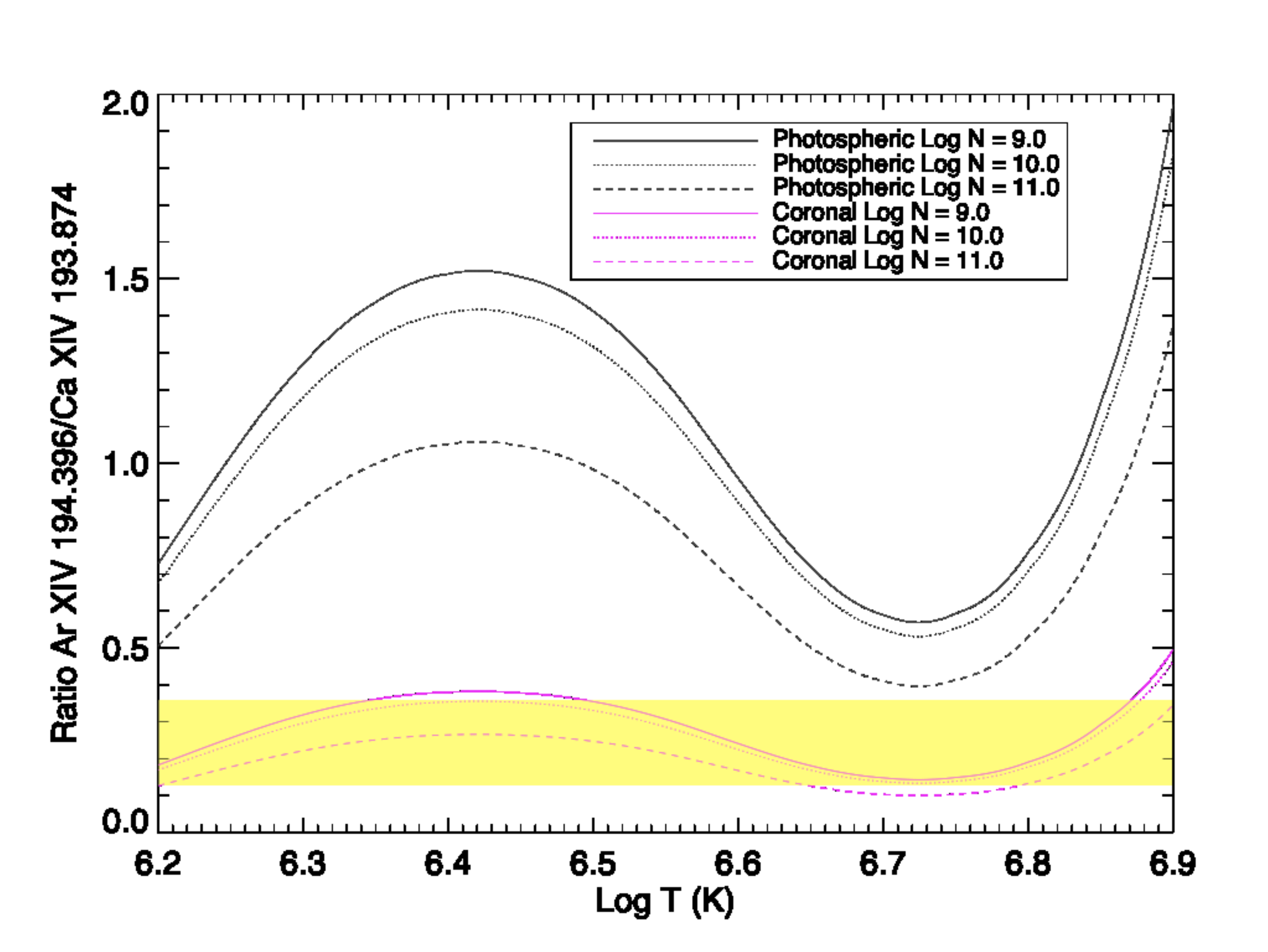}
\caption{Intensity ratios of Ar {\sc xiv} and Ca {\sc xiv} contribution functions vs electron temperature for electron densities of log$_{10}$ \emph{N} = 9.0, 10.0, 11.0. Shaded box shows the range of the Ar {\sc xiv}/Ca {\sc xiv} ratio in the temperature range  of [6.2, 6.8] for electron density of log$_{10}$ \emph{N} = 10.0 and coronal abundances as given in Section \ref{sec:ratio}.  The ratio is quite narrow for a wide temperature range but exceeds 0.50 for temperatures above log$_{10}$ \emph{T} = 6.9 K. }
\label{fig:goft}
\end{figure}


\begin{figure}[H]
\centering
\includegraphics[scale = 0.25]{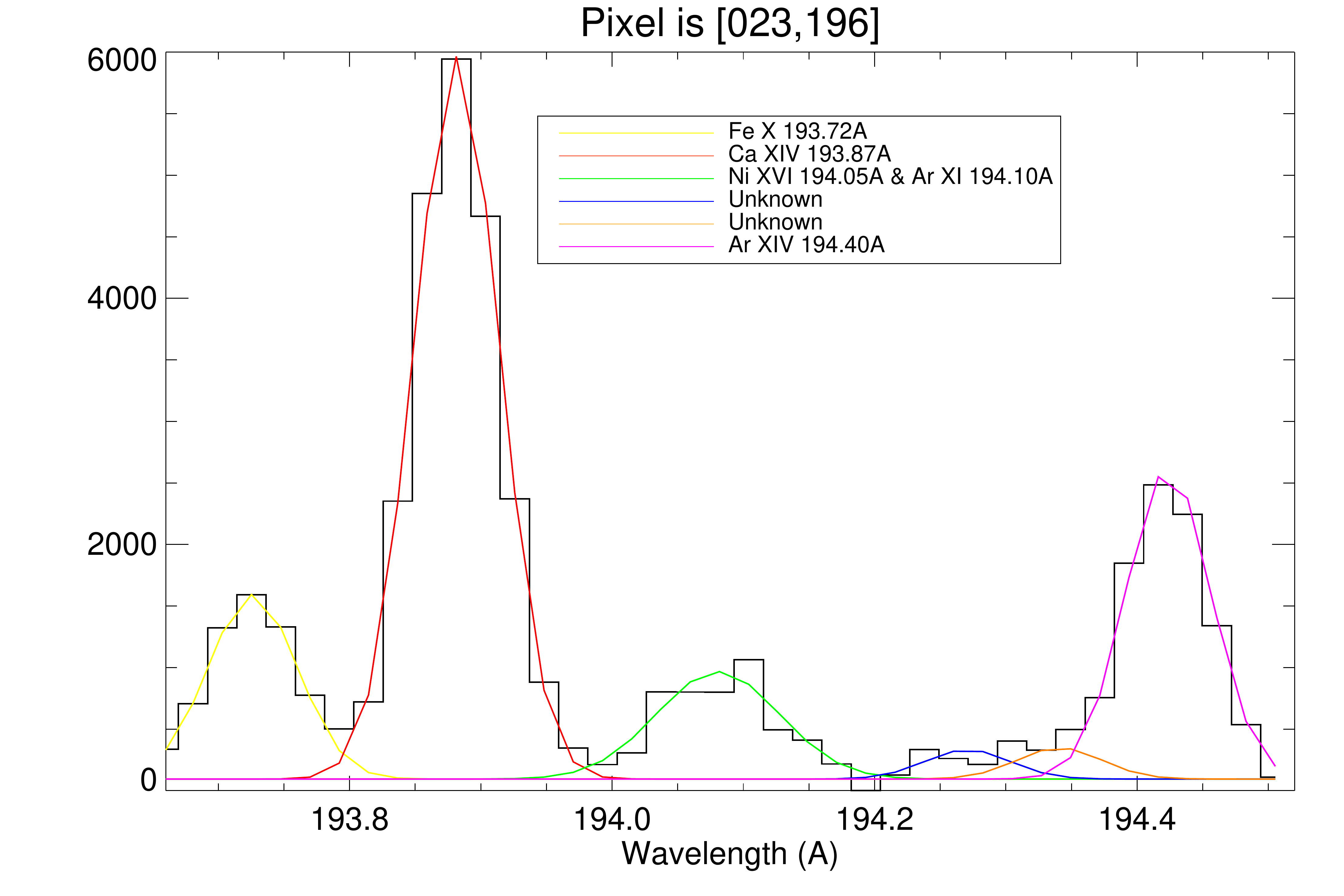}
\includegraphics[scale = 0.25]{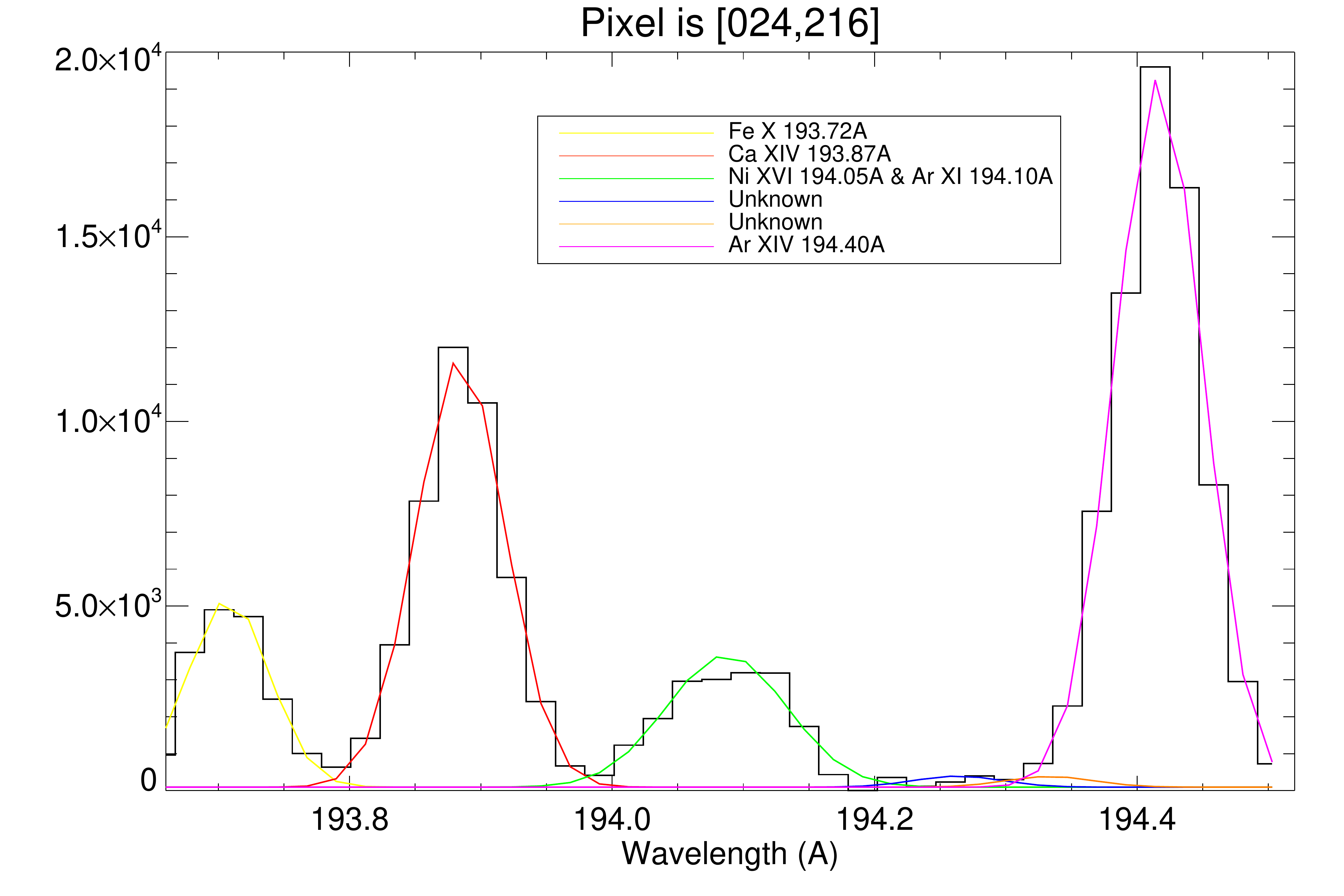}
\caption{Lines in the Ca {\sc xiv} spectral window for the \emph{Hinode}/EIS raster at 12:56 UT on March 6.  The Y-axis is intensity.  
Three Gaussians were fit to the spectral region around Ca {\sc xiv} 193.87 $\angstrom$ line (red), one for the Fe {\sc x} 193.72 $\angstrom$ (yellow) in the blue wing and the other for the Ni {\sc xvi}--Ar {\sc xi} blend (green) in the red wing.  
Three Gaussians were fit to the Ar {\sc xiv} 194.40 $\angstrom$ (magenta), two of which lie in the blue wing and are unidentified lines \citep[blue and orange;][]{brown08}. Top panel: FIP effect spectrum for pixel = [23, 196].  
The Ca {\sc xiv} 193.87 $\angstrom$ line is much more intense compared to the Ar {\sc xiv} 194.40 $\angstrom$ line and FIP bias is $\sim$0.40.  Bottom panel:  IFIP effect spectrum at pixel = [024, 216].  
The Ar {\sc xiv} line is more intense than the Ca {\sc xiv} line and the IFIP bias is $\sim$1.60.
The unidentified lines in the blue wing of the Ar {\sc xiv} line are visible in FIP effect spectra but are less evident in the IFIP spectra. 
}
 \label{fig:wavelength}
\end{figure}


\begin{figure}[H]
\centering
\includegraphics[scale = 0.60]{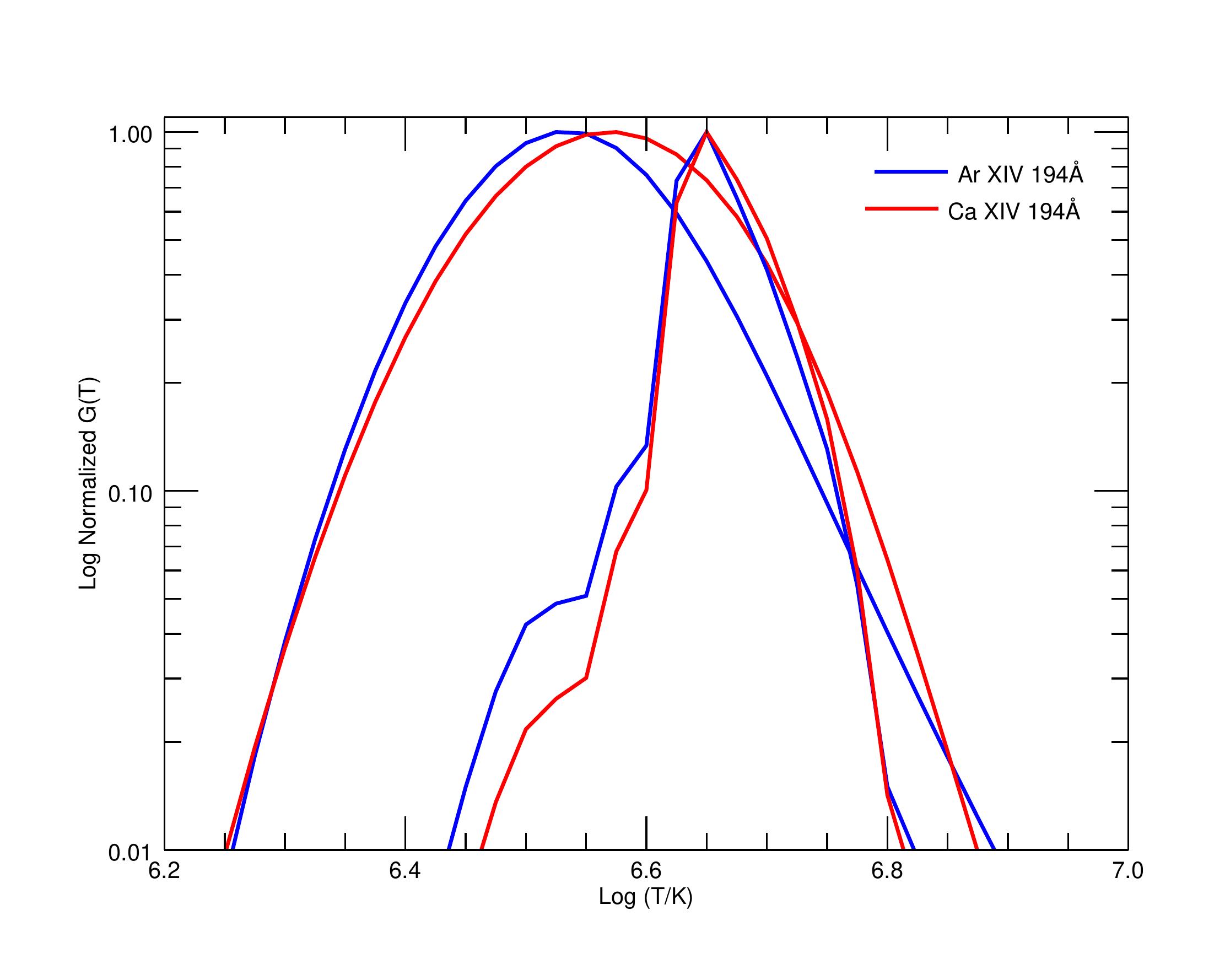}
\caption{Log normalized contribution functions, G(T), of Ar {\sc xiv} and Ca {\sc xiv} (smoothly peaked curves) and G(T) convolved with DEM (sharply peaked curves) for the IFIP patch discussed in the last paragraph of Section \ref{sec:ratio}. }
\label{fig:norm_goft}
\end{figure}

\acknowledgments
We thank the referee for their constructive comments and suggestions which have improved the manuscript.
The authors are very grateful to Konkoly Observatory, Budapest, Hungary, for hosting two workshops on Elemental Composition in Solar and Stellar Atmospheres (IFIPWS-1, 13-15 Feb, 2017 and IFIPWS-2, 27 Feb-1 Mar, 2018) and acknowledge the financial support from the Hungarian Academy of Sciences under grant NKSZ 2018$\_$2. 
The workshops have fostered collaboration by exploiting synergies in solar and stellar magnetic activity studies and exchanging experience and knowledge in both research fields.
Hinode is a Japanese mission developed and launched by ISAS/JAXA, collaborating with NAOJ as a domestic partner, and NASA and STFC (UK) as international partners. 
Scientific operation of Hinode is performed by the Hinode science team organized at ISAS/JAXA. 
This team mainly consists of scientists from institutes in the partner countries. 
Support for the post-launch operation is provided by JAXA and NAOJ (Japan), STFC (UK), NASA, ESA, and NSC (Norway). 
\emph{SDO} data were obtained courtesy of NASA/\emph{SDO} and the AIA and HMI science teams.
D.B. is funded under STFC consolidated grant number ST/N000722/1. 
L.v.D.G. is partially funded under STFC consolidated grant number ST/N000722/1. 
The work of D.H.B. was performed under contract to the Naval Research Laboratory and was funded by the NASA Hinode program. 
A.W.J., L.M.G., G.V. and L.v.D.G acknowledge the support of the Leverhulme Trust Research Project Grant 2014-051 and the Royal Society.
J.M.L. was supported by the NASA HSR (NNH16AC391) and LARS (NNH17AE601) programs, the Chandra GO program and by Basic Research Funds of the Chief of Naval Research.
D.M.L. acknowledges support from the European Commission's H2020 Program under the following Grant Agreements: GREST (no. 653982) and Pre-EST (no. 739500) as well as support from the Leverhulme Trust for an Early-Career Fellowship (ECF-2014-792) and is grateful to the Science Technology and Facilities Council for the award of an Ernest Rutherford Fellowship (ST/R003246/1).

\bibliographystyle{aasjournal}
\bibliography{abundance} 

\end{document}